\documentclass[12pt]{article}

\usepackage{amsmath}
\usepackage{graphicx}
\usepackage{overcite}
\usepackage[margin=1in]{geometry}

\makeatletter
\renewcommand{\section}{\@startsection{section}{1}{\z@}%
    {5ex \@plus -1ex \@minus -.2ex}{2ex \@plus.2ex}%
    {\normalfont\large\sffamily\bfseries\MakeUppercase}}
\renewcommand{\subsection}{\@startsection{subsection}{1}{\z@}%
    {4ex \@plus -1ex \@minus -.2ex}{1ex \@plus.2ex}%
    {\normalfont\large\sffamily\bfseries}}
\makeatother

\begin{document}

\begin{center}{\noindent\Large\sffamily\bfseries Interaction 
energies of monosubstituted benzene dimers via nonlocal density 
functional theory}\par\bigskip
T. Thonhauser, Aaron Puzder,\footnote{Present address:\it\ Lawrence 
Livermore National Laboratory, Livermore, California 94550, USA}
and David C. Langreth\\
\it Department of Physics and Astronomy, Rutgers University\\ 
Piscataway, New Jersey 08854-8019, USA
\end{center}

\noindent We present density-functional calculations for the
interaction energy of monosubstituted benzene dimers. Our approach
utilizes a recently developed fully nonlocal correlation energy
functional, which has been  applied to the pure benzene dimer and
several other systems with promising results. The interaction energy as
a function of monomer distance was calculated for four different
substituents in a sandwich and two T-shaped configurations. In
addition, we considered two methods for dealing with exchange, namely
using the revPBE generalized gradient functional as well as full
Hartree--Fock. Our results are compared with other methods, such as 
M{\o}ller--Plesset and coupled-cluster calculations,  thereby
establishing the usefulness of our approach. Since our
density-functional based method is considerably faster than other
standard methods, it provides a computational inexpensive alternative,
which is of  particular interest for larger systems where standard
calculations are too expensive or infeasible.

\newpage

\section{Introduction}
With the vast development of computational facilities in the last
decade, it is now possible to analyze small aromatic systems
numerically with high precision. Particular interest has emerged for
the non-covalent van der Waals interaction, important in many areas and
responsible for phenomena, such as DNA base pair bonding, protein
structure and folding, and organic molecule crystallization. While the
full DNA structure itself is still out of scope, calculations for the
benzene dimer, which is the simplest prototype of aromatic $\pi$--$\pi$
interaction, have been
reported.\cite{hobza96,tsuzuki92,sherrill02,sherrill04_2}  A step
toward larger systems are calculations for substituent effects on the
benzene dimer.\cite{sherrill04} State-of-the-art tools for such
calculations are second-order M{\o}ller--Plesset (MP2) perturbation 
theory and coupled-cluster calculations with singles, doubles, and
perturbative triples (CCSD(T)). These methods, especially the latter,
are regarded as highly accurate and reliable. Unfortunately, their
application to large systems is time consuming and computationally very
demanding. Furthermore, the results might depend on the basis set used
and elaborate techniques are necessary to estimate the basis set limit.

Density functional theory (DFT), especially in the form of various
generalized gradient approximations (GGA) including those with no
empirical input,\cite{PW91,PBE} have had a degree of success in
describing dense bulk matter,\cite{GGAsolid} and isolated
molecules.\cite{GGAmolecule} For molecules, this performance has been
further improved by the use of various empirical and hybrid methods.
However, for van der Waals complexes and sparse matter including many
layered structures, polymer crystals, organic molecular crystals and
the like, such methods either give sporadic results or fail
completely. In particular, the importance of DFT-based methods for
calculations of molecular clusters has been minor. The reason for this
is that the dominant part  of the stabilization energy in many cases
comes from the dispersion  energy, which is usually not included
correctly in  DFT.\cite{spooner96,pulay94, ruiz95,rydberg03,dion04}
Moreover,  functionals of the GGA type often do not even predict a
binding of  molecular dimers. Therefore, considerable effort has been
devoted to develop DFT based methods that are capable of treating van
der Waals systems. Among these are methods with empirical damping
functions,\cite{elstner01,scoles01} self-consistent charge-density
functional tight-binding methods,\cite{elstner98} and self-consistent
LCAO methods with standard pseudopotentials.\cite{sanchez97} 

Recently, a new approach using a van der Waals density functional
(vdW-DF) with a nonlocal correlation-energy has been
developed.\cite{dion04} This formalism includes van der Waals forces in
a seamless fashion and the applicability of this approach was first
demonstrated for bulk layered systems such as graphite, boron nitride,
and molybdenum sulfide.\cite{rydberg03} Similarly, it has been applied
to two-layer dimers of these same materials\cite{langrethIJQC} and the
benzene dimer in a number of geometries, where it also proved to give a
realistic description.\cite{dion04,aaron05}. More recently, it has been
applied to naphthalene, anthracene, and pyrene
dimers\cite{svetla2,svetla1} where for napthalene it gives a creditable
comparison with recent CCSD(T) results\cite{tsuzuki-naphthalene}. These
larger dimers show the power of the method for more sizable systems
where CCSD(T) results appear to be nonexistent. In addition, there have
been promising applications to a molecule physisorbed on an infinite
surface\cite{benz-graph}, as well as to a bulk crystal of
polyethylene.\cite{pe} 

In this paper, we investigate the capability of vdW-DF to describe
larger systems. We go beyond the pure benzene dimer and use the
nonlocal functional to calculate the effect of the substituents OH,
CH$_3$, F, and CN. These each replace an H atom on one of the molecules
of the benzene dimer, and among other things act as electron donors (OH
and CH$_3$) or acceptors (F and CN), as discussed by Sinnokrot and
Sherrill,\cite{sherrill04} who performed MP2 and CCSD(T) calculations.
Our results show that vdW-DF, which is much faster than MP2 and
CCSD(T), gives interaction energies and distances comparable with these
standard methods. Furthermore, it generates the correct energy ordering
of different substituent structures. Thus, it is a promising method for
larger applications such as DNA base-pair stacking and beyond, as well
as for various types of extended systems where wave function methods
are typically impractical.

This paper is organized as follows: Section~II describes the energy
functional and the details of its evaluation. In Sec.~III, we give
computational details. Our results are analyzed in Sec.~IV, where we
also compare with other numerical methods. We conclude in Sec.~V and
give an outlook for future work.

\section{Van der Waals density functional (vdW-DF)}
At the heart of every DFT calculation lies the energy functional
$E[\rho]$, which depends on the electron density $\rho$.\cite{DFT} For
our calculations we used the following particular functional with 
nonlocal correlation energy,
\begin{equation}\label{equ:E}
E[\rho]=T_s[\rho]+V_{pp}[\rho]+J[\rho]+E_x[\rho]+E_c^{\rm L}[\rho]
   + E_c^{\rm NL}[\rho]\;,
\end{equation}
which was especially designed to include the van der Waals
interaction.\cite{rydberg03,dion04} Here, $E$ is the total energy
functional of the dimer, $T_s$ is the single-particle kinetic energy,
$V_{pp}$ is the ionic pseudopotential functional, and $J$ describes the
Coulomb interaction. For the exchange functional $E_x$ we used the
revPBE parameterization of GGA.\cite{revPBE} Furthermore, $E_c^{\rm L}$
denotes the local part of the correlation, which was evaluated using
LDA\cite{LDA}. The  nonlocal part can be written as
\begin{equation}
E_c^{\rm NL}[\rho] = \frac{1}{2}\iint\! d^3r\,d^3r'\;\rho(\vec{r}\,)
\phi(\vec{r},\vec{r}\,')\rho(\vec{r}\,')\;,
\end{equation}
where $\phi(\vec{r},\vec{r}\,')$ is a function depending on
$|\vec{r}-\vec{r}\,'|$ and $\rho$ in the vicinity of $\vec{r}$ and
$\vec{r}\,'$.   Details about its construction can be found in
Ref.~[\citen{dion04}]. No empirical input was applied.

For the charge density $\rho$, we used the result of a self-consistent
DFT calculation utilizing the Perdew-Burke-Ernzerhof (PBE)
exchange-correlation functional.\cite{PBE} With this charge density we
then evaluated the functional from Eq.~(\ref{equ:E}). Differences in
interaction energies for charge densities resulting from other
exchange-correlation functionals turned out to be negligible.

\section{Computational methods}
Our self-consistent DFT calculations for obtaining the charge density 
$\rho$ were performed with the planewave code ABINIT.\cite{gonze02} We
used Troullier-Martins  pseudopotentials\cite{troullier93} with a
kinetic energy cutoff of 50 Rydberg. The dimers and monomers were
placed in boxes of such size that one system and its mirror image is
separated by 15 to 20~\AA\ of vacuum. This distance proved to
be sufficient for the charge density to approach zero well before the
supercell edge, which eliminates any interaction between periodic
replica, aside from the possibility of electrostatic as discussed
below. Even with this conservative box size, the calculations of this 
GGA part go quickly, requiring only around two hours on a single
processor (e.g. Athlon MP2000+) for each  position of the monomers
considered.

Most parts of the energy functional from Eq.~(\ref{equ:E}) were also
directly evaluated in ABINIT. Only the nonlocal correlation part was
calculated in a post-process procedure. Thereby, we computed $E_c^{\rm
NL}$ on a real space grid with a spacing equal to the Fourier transform
spacing in the plane wave calculations, and then we performed a
numerical integration as described in Ref.~[\citen{dion04}]. This
procedure, as currently coded, requires a computer time comparable to
the DFT calculation. Although optimization of the code could
substantially cut this time, that effort is probably unnecessary, as
the time for the straightforward GGA calculation of the above paragraph
will dominate as the system size increases.

At this point we should mention another promising complementary  method
(SAPT-DFT),\cite{misquitta,szal1} a density functional based version of
symmetry adapted perturbation theory (SAPT), which  uses time-dependent
DFT to calculate the dispersion contribution. This complementary
procedure appears to have advantages and disadvantages with respect to
our much simpler one. First, it  gives\cite{szal1} an energy curve
closer to CCSD(T) than vdW-DF for the benzene dimer\cite{aaron05} in
the sandwich geometry and also for rare gas dimers. Furthermore, with 
recent optimization, the computer time required for benzene is (like
vdW-DF)  comparable to that of an ordinary density functional
calculation.\cite{szal2} It appears likely that for systems this size
or a little larger, this better accuracy may continue.  However, the
computer time of SAPT-DFT scales as the fifth power of the basis size,
so that its application will rapidly become impractical as the system
size increases, as opposed to vdW-DF, which will continue to scale as
an ordinary DFT calculation.  In addition, SAPT-DFT is not a density
functional; its application necessitates the identification of
individual fragments, so it cannot be seamless as fragments merge
together to become single entities. It is also not obvious how it might
be applied to a single fragment for which dispersion interaction
between different ends or parts is important; nor to an extended system
that cannot be divided into individual summable finite parts. Being a
simple density functional, vdW-DF can do all of the  above. In summary,
it can be applied to calculations of all types and sizes for which an
ordinary LDA or GGA calculation can.  These complementary aspects of
SAPT-DFT and vdW-DF render both methods important to pursue.

Because the GGA part of our calculation is performed on a system
consisting of replicas of the relevant molecular complex separated by
40 Bohr on a simple cubic lattice, one must consider the possibility of
spurious electrostatic interactions between these replicas showing up
in our results. Of course, the pure benzene monomer does not have a
permanent dipole moment. On the other hand, the dipole moments of the
substituted  benzene range from $\sim$0.4 debye for toluene to $\sim$4
debye for benzonitrile.  The Coulomb energy associated with this
interaction is easily calculated using a method attributed to 
Lorentz\cite{lorentz} (see Jackson\cite{jackson} for a more easily
accessible description). We find that this Coulomb energy ranges from
$\sim$0.05 kcal/mol for benzonitrile and becomes two orders of
magnitude smaller for toluene.  The former number is comparable to the
numerical accuracy of our calculations.  We cite it because it is a
hard number that must reasonably be expected to be larger than any
induced electrostatic interactions, which we will estimate below. This
direct interaction obviously cancels out of our calculation exactly
when the subtraction is made to find the energy of interaction between
the two parts of each molecular complex.  However, the  dipole moment
of the complex in one supercell can induce a changed charge
distribution in each of the two molecules in an adjacent supercell,
thereby changing the electrostatic interaction between the latter. 
This interaction does not cancel out, but can be estimated from the
dipole moments and polarizabilities involved.  We find that it gives a
contribution ranging from $\sim$$10^{-4}$ kcal/mol for the CH$_3$
substituted dimer to $\sim$$10^{-2}$ kcal/mol for the CN substituted
one.  On the level of the other inaccuracies in our calculation, these
errors are completely negligible.

Of the exchange functionals tried
earlier\cite{rydberg03,langrethIJQC,dion04}, Zhang and Yang's
``revPBE''  was the closest to Hartree--Fock (HF) for the dimers
compared\cite{langrethIJQC,dion04}.  However, it was clear that the
magnitude of  its slope was a little too big in the binding region.
This lead to the speculation that the equilibrium distances  with
revPBE exchange, which were also a little too big,  would be improved
if full HF exchange were used instead. This conjecture was verified in
Ref.~[\citen{aaron05}],  where the use of HF exchange instead of revPBE
exchange in the functional of Eq.~(\ref{equ:E}) was fully implemented
and applied to the benzene dimer in several different configurations.
There, the exchange $E_x$ was evaluated both with the revPBE exchange
functional and HF, and the various tradeoffs evaluated and discussed.
Here, we do the same for the substituted dimers. For HF exchange we
used the general ab-initio  package GAMESS,\cite{gamess} using a triple
zeta valence (TZV) basis set.

\section{Results and Discussion}
We calculated interaction energies for three typical configurations of
the dimer,\cite{hobza96,tsuzuki92,sherrill02,sherrill04_2} as depicted
in Fig.~1. In the sandwich structure, the two monomers are arranged
parallel in such a way that the centers of the rings are on top of each
other. In the T-shaped structure, the two monomers form an angle of
90$^\circ$, where one monomer is placed right above the center of the
other ring. We have chosen a configuration in which the substituent
points away from the other ring [T-shaped(1)] and a similar
configuration in which both monomers are rotated by 90$^\circ$
[T-shaped(2)]. For our calculations we have considered four different
substituted molecules, namely phenol, toluene, fluorobenzene, and
benzonitrile (whose substituents are OH, CH$_3$, F, and CN,
respectively). In addition, we also calculated the pure benzene dimer,
whose ``substituent'' we refer to as ``H''. In all cases, we  define
``distance'' as the distance from the center of the benzene ring to the
center of the aromatic ring containing the substituent. All atomic
positions for the monomers were optimized during the self-consistent
DFT calculations, however, for the calculations of interaction
energies, the two constituent monomers were considered ``rigid'' and
only the intermonomer distance was changed.

\subsection{Sandwich configuration}
Results for the interaction energy as a function of the distance for
the sandwich configuration are plotted in Fig.~2.  Each line depicted
consists of 20 points for distances from 3 to 8~\AA. The data was then
fitted with splines to be continuous. As expected, for short distances
all curves show strong repulsion, whereas for large distances the
interaction vanishes. Furthermore, all curves show a pronounced minimum
around 4\AA. Values for the interaction energy at these minima can be
found in Table~\ref{tab:energies}, where we also compare with other
numerical methods. Unfortunately, corresponding experimental values are
not available, however, we want to mention that some experimental
studies of substituent effects in related systems are 
reported,\cite{paliwal94,rashkin02} but it is difficult to reliably
extract correct binding energies.

It can be seen in Table~\ref{tab:energies} that the values for the
binding distances are approximately 10\% larger than the corresponding
MP2 values and about 5\% larger than CCSD(T) values. Numbers for the
interaction energies are in between MP2 and CCSD(T) values, and about 
25\% larger compared to CCSD(T). This can be partly explained by the
fact that DFT calculations underestimate the quadrupole moment by as
much as 20\%.\cite{quad} This makes our interaction energies too large,
since not enough repulsion is considered.

In a simple picture, as pointed out by Sinnokrot and 
Sherill,\cite{sherrill04} one could think of the substituents as
electron donors (OH, CH$_3$) or electron withdrawing (F, CN). In terms
of a Hunter--Sanders model,\cite{hunter90} where electrostatics is
assumed to determine changes in binding, one would expect to find that
dimers with F and CN substituents bind more strongly than the pure
dimer, whereas OH and CH$_3$ bind more weakly.  On the contrary, they
found increased binding for all the substituted species. A subsequent
breakdown\cite{sherrill04} using SAPT with a modest basis set showed
these differences  to be distributed among a number of different energy
components, whose individual values were rather sensitive to the
precise values of intermonomer distance.   We might expect the binding
from dispersion to increase due to the addition of substituents, as
they would be expected to reduce the energy gap for electronic
particle-hole excitations, and hence increase the polarizability of the
substituted monomer. However, SAPT calculations showed this effect to
be too small to alone explain the differences. Nevertheless, it was
clear that dispersion was by far the largest individual attractive
component of the energy, and the necessity to be able to calculate this
with little variation in fractional error for each substituted species
is crucial to obtain a correct overall picture.

In the light of the above, it is an important challenge to the vdW-DF
density functional to see if it can reproduce the changes that occur
upon substitution, and hence it is interesting to look at our predicted
interaction energies relative to  those of the pure benzene dimer.
Numbers for these values at the corresponding minimum are collected in
Table~\ref{tab:energy_diff}. If we also normalize the distance such
that it is measured relative to the benzene binding distance, the plot
in the right panel of Fig.~2 is obtained. Here, we also added the
CCSD(T) minimum points so that a direct comparison is possible. It can
be seen clearly that the DFT method is in very good agreement with the
CCSD(T) calculations. Similarly, we predict a bond length contraction
of $\sim$0.1 \AA\ for all the substituents, in agreement with the
CCSD(T) calculations. In general, while all interaction energies depend
somewhat upon the specific details of the calculation, the relative
energies for the  different substituents are almost independent
thereof.

\subsection{T-shaped(1) and T-shaped(2) configuration}
Results for interaction energies of T-shaped(1) and T-shaped(2) dimers
are plotted in Figs.~3 and 4. All curves show pronounced minima around
5.25~\AA. Values for $E_\text{int}$ at the minimum points and relative
to the benzene dimer can again be found in Tables~\ref{tab:energies}
and \ref{tab:energy_diff}, respectively. It can be seen immediately
that the  effect of substitution in the T-shaped configurations is much
weaker. The difference in $E_\text{int}$ for  the strongest and weakest
binding substituents is only approximately half of what it is in
sandwich configuration. 

As mentioned before, the quadrupole-quadrupole interaction within DFT
is underestimated. For sandwich structures, this leads to a binding
that is too strong, since not enough repulsion is considered. For
T-shaped configurations, however, the quadrupole-quadrupole interaction
constitutes an attraction, since the monomers form an angle of
90$^\circ$. Here, the underestimation will lead to a binding which is
too weak. The effect will be smaller though, because the T-shaped
binding distances are larger than the sandwich ones. As a result, the
vdW-DF interaction energies now show a weaker binding compared to MP2
and CCSD(T). The binding order, on the other hand, is in perfect
agreement. In particular, the isoenergetic levels of H and OH in
T-shaped(1) and T-shaped(2) configuration are well reproduced. 

As for the sandwich structure, we calculated the interaction energies
in reference to the benzene dimer and its binding distance. The results
are depicted in the right graphs of Figs.~3 and 4. Again, it can be
seen that the DFT method is generally in quite good agreement with the
CCSD(T) calculations.

\subsection{Hartree--Fock exchange}
As pointed out in Ref.~[\citen{aaron05}], using HF exchange instead of
revPBE while evaluating the functional of Eq.~(\ref{equ:E}) influences
binding distances and interaction energies.  While the exchange energy
might alter the shape of the potential curve to a smaller or greater
extent, it is the nonlocal correlation energy functional that allows to
treat van der Waals systems within a DFT approach. As we will see
below, the binding order of different substituents is actually
independent of the exchange functional used and thus, it is governed by
the nonlocal correlation energy  and other components of the total
energy. From this point of view, it seems conceivable to utilize even
other exchange functionals than revPBE or HF. However, there is one
important point to keep in mind: Within a given covalently bonded
monomer the full HF interaction can be long ranged, because the Coulomb
interaction can extend from one end of a molecular orbital to another,
whereas the proper inclusion of correlation will shield this
interaction, yielding an effectively short-ranged Coulomb interaction
(aside from van der Waals effects). In ordinary DFT (either LDA or GGA)
this is handled by making both exchange and correlation short-ranged,
thus mixing their definitions in this respect. 
 
We repeated all calculations from above using HF exchange instead of
revPBE. The results for all configurations are plotted in Fig.~5. In
general, the effect of using HF exchange instead of revPBE is a
stronger binding at shorter distances. Values for the interaction
energy at the minimum points can be found in Tables~\ref{tab:energies}
and \ref{tab:energy_diff}, where we refer to the DFT calculations with
HF exchange as vdW-DF(HF). Remarkably, the relative order of
interaction energies for different substituents is  the same with HF
and revPBE exchange. It can also be seen that the effect of using HF
exchange instead of revPBE is stronger in the T-shaped configuration
than in the sandwich structure. However, it is clear that the  energy
differences between the benzene dimer and the four substituted ones are
not in such good agreement with CCSD(T) as the results using revPBE
were. As the substituted monomers are significantly longer  and
presumably more easily polarized than benzene alone, one can speculate
that this poorer agreement is due to the lack of long range shielding
of the Coulomb potential in the HF approximation, as discussed in the
above paragraph.  Alternatively, it could be due to the lack of
self-consistency of  simply substituting the self-consistent HF
exchange energy for the self-consistent GGA one, which is a related
problem. In any case, the larger issue of the consistency of various
exchange approximations with the correlation method used is an
important one, and a subject of further study.

However, with respect to the grosser effects of HF exchange, we find
results similar to Ref.~[\citen{aaron05}], in particular an improvement
in the binding distances. The vdW-DF(HF) values are now within 
$\sim$1\% of the CCSD(T) values. On the other hand, the interaction
energies now show a stronger binding than the vdW-DF results. The use
of HF exchange also corrects the relation of interaction energies
between T-shaped and sandwich structures: While vdW-DF always predicts
the  sandwich structure to be lower in energy, vdW-DF(HF)  reverses 
this order and gives stronger binding for all T-shaped structures, in
agreement with CCSD(T) calculations.\cite{sherrill04}

\section{Conclusions}
We have presented DFT-based calculations for systems that exhibit van
der Waals type bonding. In particular, we investigated substituent
effects on benzene dimers for several different substituents. Our
method gives useful approximations for interaction energies and binding
distances, when compared with state-of-the-art wave function methods,
which are computationally much more demanding.  In addition, the
changes produced by the substituents as given by our vdW-DF method were
quite close indeed to those predicted by CCSD(T). Our DFT based method
offers a computationally inexpensive alternative that can speed up
calculations drastically, and in addition make applications to extended
systems possible.  To be specific, our calculations require only
slightly more computer time for large systems than the GGA type of DFT
calculations that they replace. With this in mind, it is a promising
candidate for larger applications, such as  DNA base-pair stacking,
which is already work in progress.  The method is also applicable to
extended systems where wave function methods are infeasible.

\section*{Acknowledgments}
We thank Maxime Dion, as well as the Chalmers members of the
Chalmers-Rutgers collaboration, especially Bengt Lundqvist, for helpful
communications and discussions. This work was supported by NSF Grant
No.\ DMR-0456937. All calculations were performed on the Rutgers
high-performance supercomputer facility, operated by the Center for
Materials Theory of the Department of Physics and Astronomy.

\setlength{\parindent}{0mm}

\newpage

\begin{table}\renewcommand{\arraystretch}{0.8}
\caption{\label{tab:energies} Comparison of interaction energies
$E_\text{int}$ and  binding distances $R$ for all substituents,
calculated with different methods. MP2 and CCSD(T) values are taken
from Ref.~[\citen{sherrill04}], where calculations were performed with
an aug-cc-pVTZ basis set.}
\begin{tabular*}{\textwidth}
{@{}l@{\extracolsep\fill}lc@{}cc@{}cc@{}c@{}}
\hline\hline
& & \multicolumn{2}{c}{sandwich} & 
    \multicolumn{2}{c}{T-shaped(1)} & 
    \multicolumn{2}{c}{T-shaped(2)}\\
  & method & $R$ [\AA] & $E$ [kcal/mol] & $R$ [\AA] & $E$ [kcal/mol] 
           & $R$ [\AA] & $E$ [kcal/mol]\\\hline
H   &   MP2             & 3.70 & $-3.26$ & 4.89 & $-3.46$ & 4.89 & $-3.46$\\
    &   estd. CCSD(T)   & 3.90 & $-1.80$ & 4.99 & $-2.62$ & 4.99 & $-2.62$\\
    &   vdW-DF          & 4.10 & $-2.37$ & 5.28 & $-2.28$ & 5.28 & $-2.28$\\
    &   vdW-DF(HF)      & 3.83 & $-2.93$ & 4.90 & $-4.40$ & 4.90 & $-4.40$\\
OH  &   MP2             & 3.60 & $-3.75$ & 4.90 & $-3.42$ & 4.90 & $-3.52$\\
    &   estd. CCSD(T)   & 3.80 & $-2.17$ & 5.00 & $-2.58$ & 5.00 & $-2.67$\\
    &   vdW-DF          & 4.03 & $-2.79$ & 5.27 & $-2.28$ & 5.26 & $-2.41$\\
    &   vdW-DF(HF)      & 3.74 & $-3.85$ & 4.88 & $-4.45$ & 4.90 & $-4.39$\\
CH$_3$& MP2             & 3.65 & $-3.96$ & 4.90 & $-3.39$ & 4.80 & $-3.89$\\
    &   estd. CCSD(T)   & 3.80 & $-2.27$ & 5.00 & $-2.55$ & 5.00 & $-2.95$\\
    &   vdW-DF          & 4.04 & $-2.95$ & 5.25 & $-2.27$ & 5.24 & $-2.62$\\
    &   vdW-DF(HF)      & 3.75 & $-3.81$ & 4.87 & $-4.26$ & 4.88 & $-4.82$\\
F   &   MP2             & 3.70 & $-3.81$ & 4.90 & $-3.61$ & 4.90 & $-3.17$\\
    &   estd. CCSD(T)   & 3.80 & $-2.29$ & 5.00 & $-2.77$ & 5.00 & $-2.38$\\
    &   vdW-DF          & 4.03 & $-2.85$ & 5.28 & $-2.46$ & 5.29 & $-2.23$\\
    &   vdW-DF(HF)      & 3.75 & $-3.98$ & 4.89 & $-4.81$ & 4.91 & $-4.05$\\
CN  &   MP2             & 3.60 & $-4.86$ & 4.80 & $-4.11$ & 4.90 & $-3.08$\\
    &   estd. CCSD(T)   & 3.80 & $-3.05$ & 4.90 & $-3.25$ & 5.00 & $-2.20$\\
    &   vdW-DF          & 3.97 & $-3.68$ & 5.22 & $-2.86$ & 5.31 & $-2.01$\\
    &   vdW-DF(HF)      & 3.70 & $-5.16$ & 4.86 & $-5.37$ & 4.92 & $-3.85$\\
\hline\hline
\end{tabular*}
\end{table}

\newpage

\begin{table}\renewcommand{\arraystretch}{0.8}
\caption{\label{tab:energy_diff} Interaction energies relative to the
benzene dimer in [kcal/mol]. MP2/aug-cc-pVTZ and CCSD(T)/aug-cc-pVTZ
values are taken from Ref.~[\citen{sherrill04}].}
\begin{tabular*}{\textwidth}
{@{}l@{\extracolsep\fill}rrrr@{}}\hline\hline
& OH& CH$_3$ & F & CN\\\hline
\multicolumn{5}{c}{sandwich structure}\\
MP2             & $-0.49$ & $-0.70$ & $-0.55$ & $-1.60$\\
estd. CCSD(T)   & $-0.37$ & $-0.47$ & $-0.49$ & $-1.25$\\
vdW-DF          & $-0.42$ & $-0.58$ & $-0.48$ & $-1.31$\\
vdW-DF(HF)      & $-0.92$ & $-0.88$ & $-1.05$ & $-2.23$\\\hline
\multicolumn{5}{c}{T-shaped(1) structure}\\
MP2             &  $0.04$ &  0.07 & $-0.15$ & $-0.65$\\
estd. CCSD(T)   &  $0.04$ &  0.07 & $-0.15$ & $-0.63$\\
vdW-DF          &  $0.00$ &  0.01 & $-0.18$ & $-0.58$\\
vdW-DF(HF)      &  $-0.05$&  0.14 & $-0.41$ & $-0.97$\\\hline
\multicolumn{5}{c}{T-shaped(2) structure}\\
MP2             & $-0.06$ & $-0.44$ &  0.29 &  0.39\\
estd. CCSD(T)   & $-0.05$ & $-0.33$ &  0.24 &  0.42\\
vdW-DF          & $-0.13$ & $-0.34$ &  0.05 &  0.27\\
vdW-DF(HF)      &   0.01  & $-0.42$ &  0.35 &  0.55\\
\hline\hline
\end{tabular*}
\end{table}

\clearpage
\newpage\begin{quote} 

\vspace*{\fill}
FIG. 1. Models of the dimer, where we have used a CH$_3$ substituent as
example in {\bf a)} sandwich configuration, {\bf b)} T-shaped(1)
configuration, and {\bf c)} T-shaped(2) configuration. The distance is
measured between the centers of the rings.
\vfill

FIG. 2. Results for the interaction energy $E_{\text{int}}$ as a
function of the monomer separation in sandwich configuration. In the
left panel, the absolute energy values are plotted over the absolute
distance. In the right panel, the energy and distance are plotted  with
respect to the minimum energy and distance of the benzene dimer. In
addition, the CCSD(T) minimum points are depicted as corresponding 
larger symbols.
\vfill

FIG. 3. Results for the interaction energy $E_{\text{int}}$ as a
function of the monomer separation in T-shaped(1) configuration. 
{\bf(left)} The absolute energy and distance are plotted. {\bf(right)}
Energy and distance are given in reference to the minimum energy and
distance of the benzene dimer. In addition, the CCSD(T) results are
depicted.
\vfill

\newpage
\vspace*{\fill}
FIG. 4. Results for the interaction energy $E_{\text{int}}$ as a
function of monomer separation in T-shaped(2) configuration.
{\bf(left)} The absolute energy and distance are plotted. {\bf(right)}
Energy and distance are given in reference to the minimum energy and
distance of the benzene dimer. The CCSD(T) minimum points are also
depicted as corresponding symbols.
\vfill

FIG. 5. Interaction energy $E_{\text{int}}$ as a function of the
distance for sandwich, T-shaped(1), and T-shaped(2) configuration, 
where Hartree-Fock exchange was used.
\vfill
\end{quote}

\clearpage
\newpage
\pagestyle{empty}
\vspace*{\fill}
\begin{center}
\includegraphics[width=10cm]{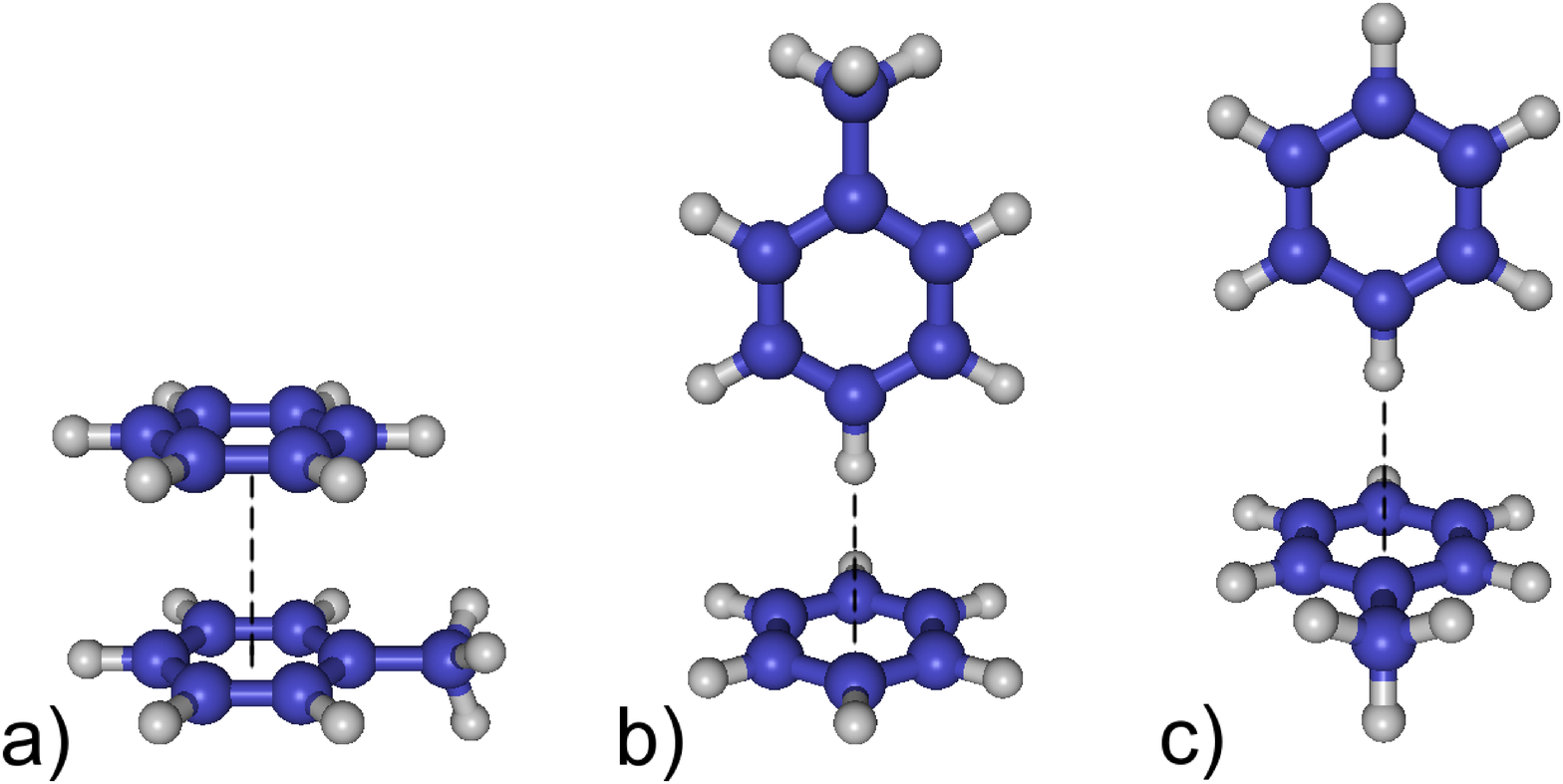}
\end{center}
\vfill
T. Thonhauser \emph{et al.}, Fig. 1

\newpage
\vspace*{\fill}
\begin{center}
\includegraphics[width=14cm]{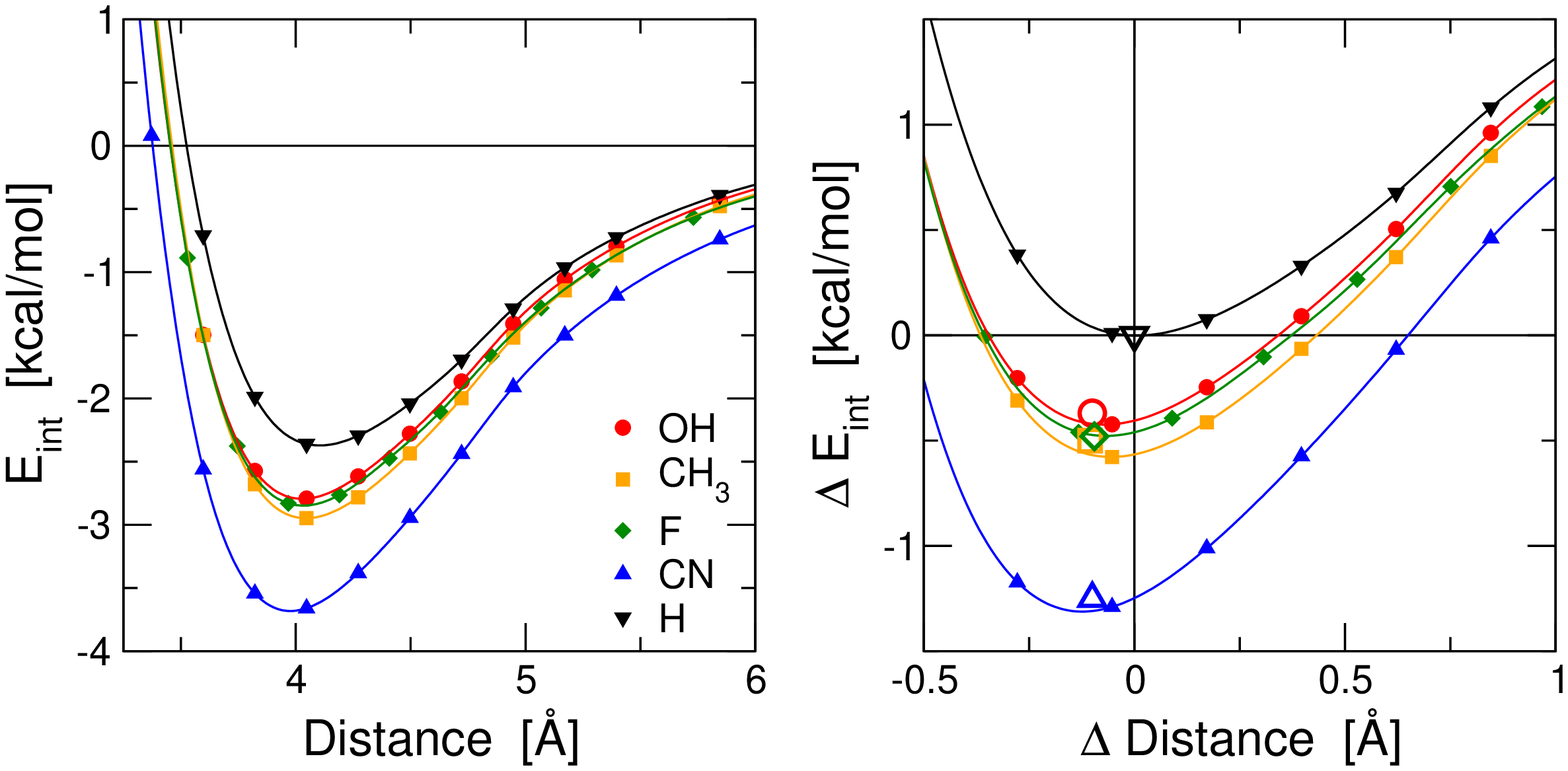}
\end{center}
\vfill
T. Thonhauser \emph{et al.}, Fig. 2

\newpage
\vspace*{\fill}
\begin{center}
\includegraphics[width=14cm]{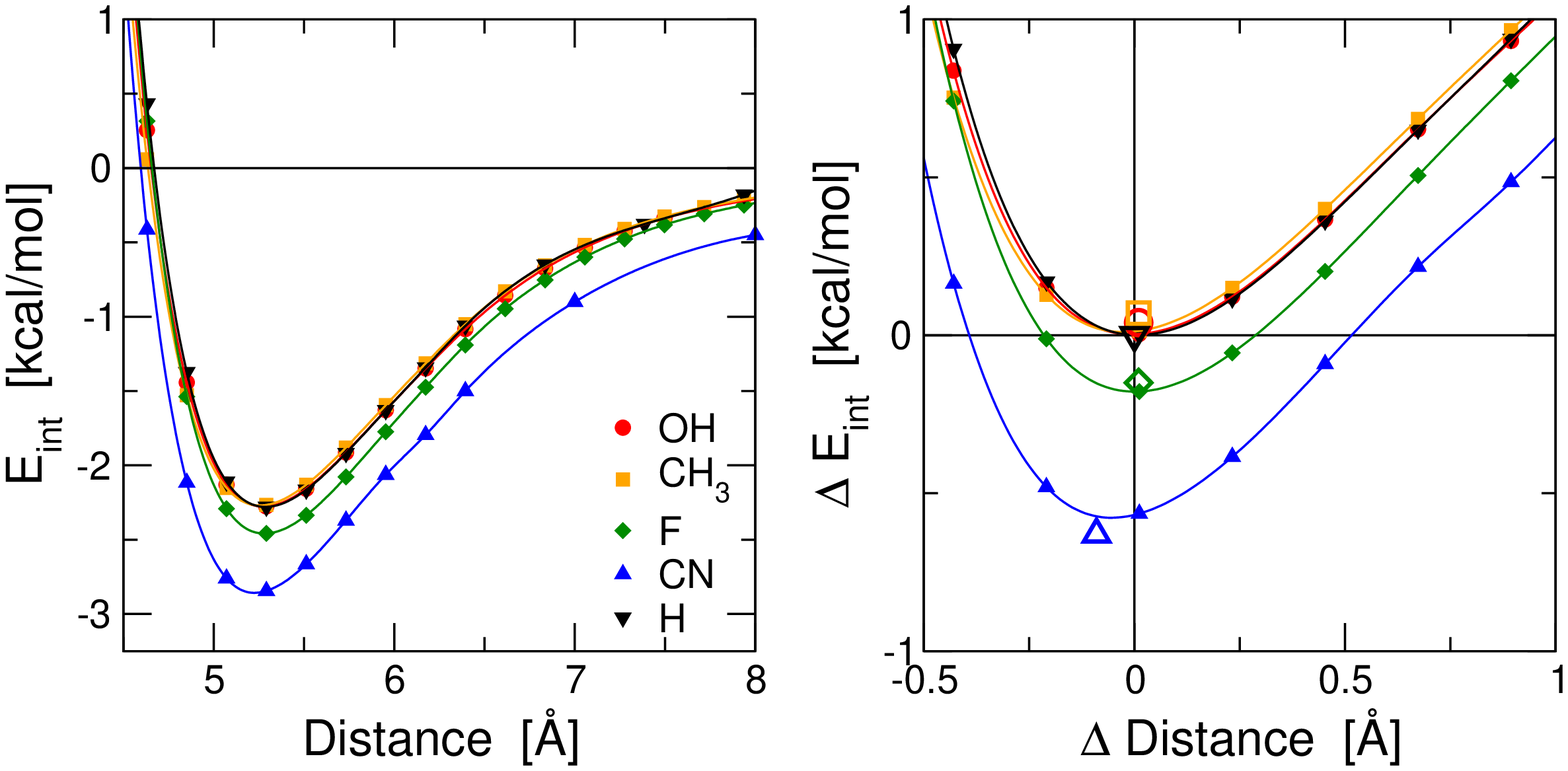}
\end{center}
\vfill
T. Thonhauser \emph{et al.}, Fig. 3

\newpage
\vspace*{\fill}
\begin{center}
\includegraphics[width=14cm]{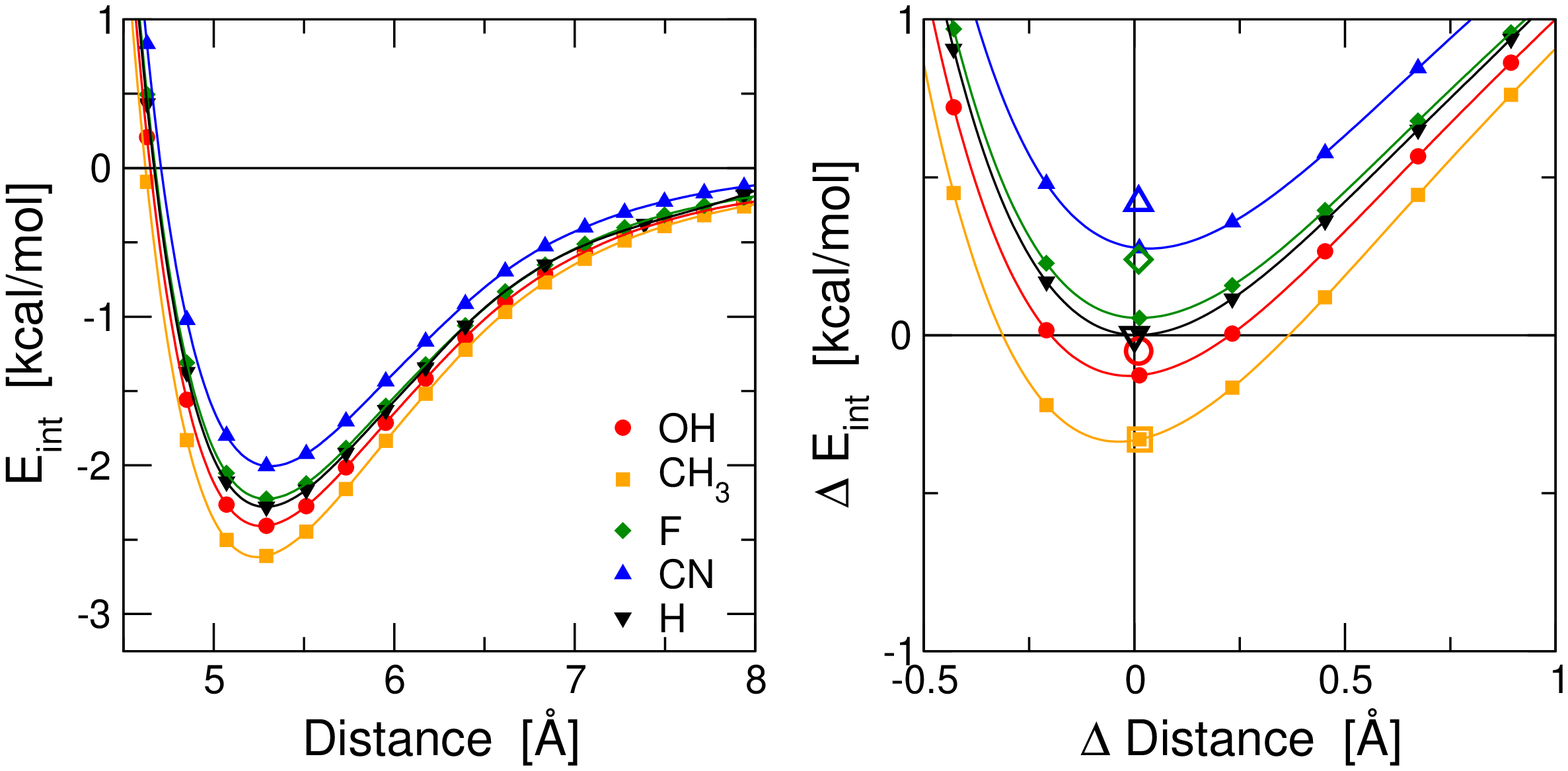}
\end{center}
\vfill
T. Thonhauser \emph{et al.}, Fig. 4

\newpage
\vspace*{\fill}
\begin{center}
\includegraphics[width=9cm]{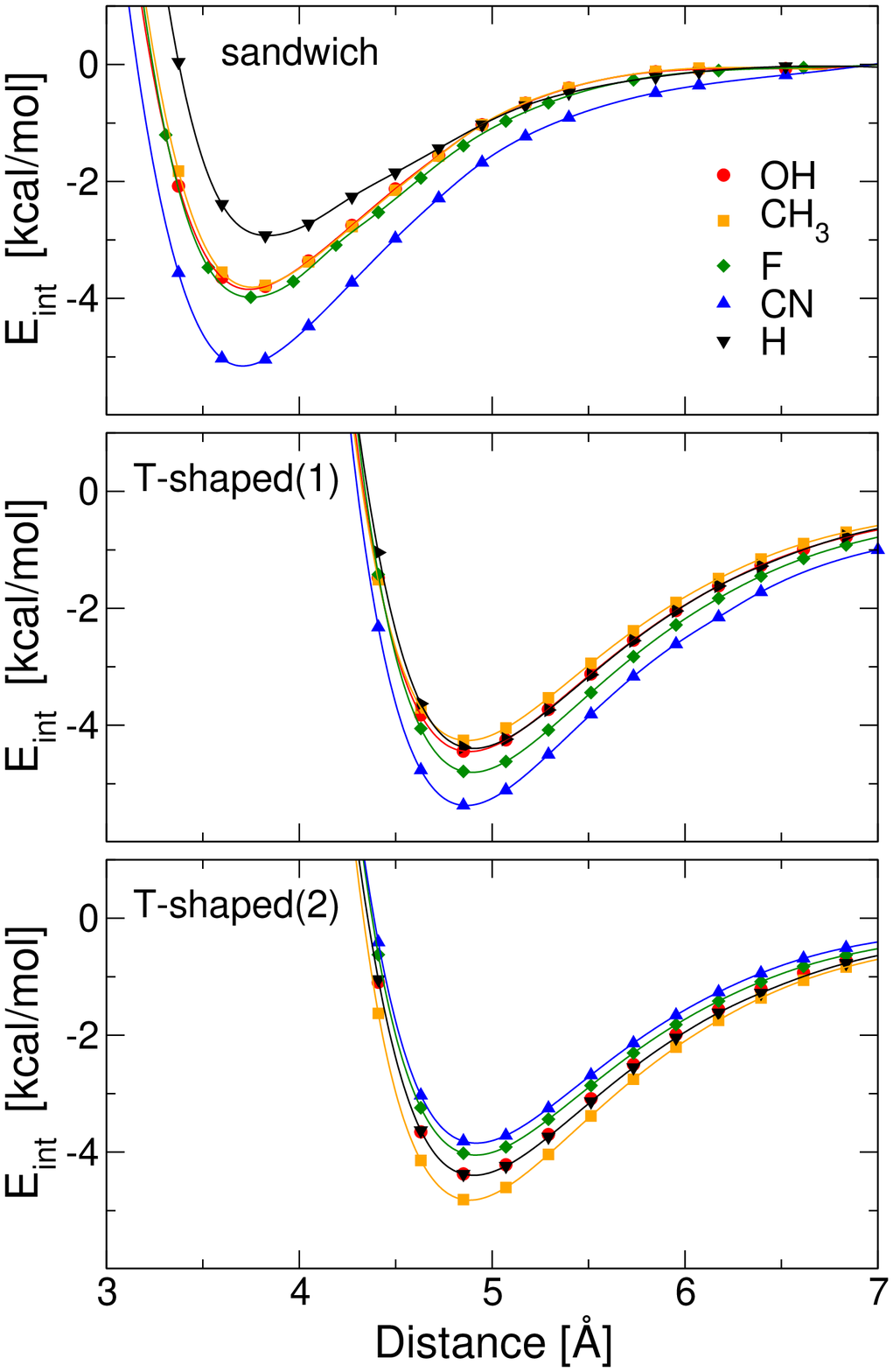}
\end{center}
\vfill
T. Thonhauser \emph{et al.}, Fig. 5


\newpage

\begin{thebibliography}{37}
\expandafter\ifx\csname natexlab\endcsname\relax\def\natexlab#1{#1}\fi
\expandafter\ifx\csname bibnamefont\endcsname\relax
\def\bibnamefont#1{#1}\fi
\expandafter\ifx\csname bibfnamefont\endcsname\relax
\def\bibfnamefont#1{#1}\fi
\expandafter\ifx\csname citenamefont\endcsname\relax
\def\citenamefont#1{#1}\fi
\expandafter\ifx\csname url\endcsname\relax
\def\url#1{\texttt{#1}}\fi
\expandafter\ifx\csname urlprefix\endcsname\relax\def\urlprefix{URL }\fi
\providecommand{\bibinfo}[2]{#2}
\providecommand{\eprint}[2][]{\url{#2}}

\bibitem{hobza96} P. Hobza, H. L. Selzle, and E. W. Schlag,
    J. Phys. Chem. {\bf 100}, 18790 (1996).

\bibitem{tsuzuki92} S. Tsuzuki, K. Honda, T. Uchimaru, M. Mikami,
    and K. Tanabe, J. Am. Chem. Soc. {\bf 124}, 104 (2002).

\bibitem{sherrill02} M. O. Sinnokrot, E. F. Valeev, and C. D. Sherrill,
    J. Am. Chem. Soc. {\bf 124}, 10887 (2002).

\bibitem{sherrill04_2} M. O. Sinnokrot and C. D. Sherrill,
    J. Phys. Chem. A {\bf 108}, 10200 (2004).

\bibitem{sherrill04} M. O. Sinnokrot and C. D. Sherrill,
    J. Am. Chem. Soc. {\bf 126}, 7690 (2004).

\bibitem{PW91}
J. P. Perdew and  J. A. Chevary and S. H. Vosko and
	Koblar A. Jackson and Mark R. Pederson and D. J. Singh,
	Phys.\ Rev.~B \textbf{46}, 6681 (1992); Erratum: {\bf 48}, 4979 (1993).

\bibitem{PBE} J. P. Perdew, K. Burke, and M. Ernzerhof,
    Phys. Rev. Lett. {\bf 77}, 3865 (1996).

\bibitem{GGAsolid}
\bibinfo{author}{\bibfnamefont{V.~N.} \bibnamefont{Staroverov}},
\bibinfo{author}{\bibfnamefont{G.~E.} \bibnamefont{Scuseria}},
\bibinfo{author}{\bibfnamefont{J.}~\bibnamefont{Tao}}, \bibnamefont{and}
\bibinfo{author}{\bibfnamefont{J.~P.} \bibnamefont{Perdew}},
\bibinfo{journal}{Phys.\ Rev.\ B} \textbf{\bibinfo{volume}{69}},
\bibinfo{pages}{075102} (\bibinfo{year}{2004}).

\bibitem{GGAmolecule}
\bibinfo{author}{\bibfnamefont{V.~N.} \bibnamefont{Staroverov}},
\bibinfo{author}{\bibfnamefont{G.~E.} \bibnamefont{Scuseria}},
\bibinfo{author}{\bibfnamefont{J.}~\bibnamefont{Tao}}, \bibnamefont{and}
\bibinfo{author}{\bibfnamefont{J.~P.} \bibnamefont{Perdew}},
\bibinfo{journal}{J. Chem.\ Phys.} \textbf{\bibinfo{volume}{119}},
\bibinfo{pages}{12129} (\bibinfo{year}{2003}).

\bibitem{dion04} M. Dion, H. Rydberg, E. Schr\"oder, D. C. Langreth,
    and B. I. Lundqvist, Phys.\ Rev.\ Lett.~{\bf 92}, 246401 (2004).

\bibitem{spooner96} J. \v{S}poner, J. Leszczynski, and P. Hobza,
    J. Comput. Chem. {\bf 17}, 841 (1996); P. Hobza, J. \v{S}poner, 
    and T. Reschel, J. Comput. Chem. {\bf 16}, 1315 (1995).

\bibitem{pulay94} S. Kristyan and P. Pulay, Chem. Phys. Lett.
   {\bf 229}, 175 (1994). 

\bibitem{ruiz95} E. Ruiz, D. R. Salahub, A. Vela, 
    J. Am. Chem. Soc. {\bf 117}, 1141 (1995).

\bibitem{rydberg03} H. Rydberg, M. Dion, N. Jacobson, E. Schr\"oder,
    P. Hyldgaard, S. I. Simak, D. C. Langreth, and B. I. Lundqvist,
    Phys. Rev. Lett. {\bf 91}, 126402 (2003).
    
\bibitem{elstner01} M. Elstner, P. Hobza, T. Frauenheim, S. Suhai,
    and E. Kaxiras, J. Chem. Phys. {\bf 114}, 5149 (2001).

\bibitem{scoles01} X. Wu, M. C. Vargas, S. Nayak, V. Lotrich, and 
    G. Scoles, J. Chem. Phys. {\bf 115}, 8748, (2001).
    
\bibitem{elstner98} M. Elstner, D. Porezag, G. Jungnickel, J. Elstner,
    M. Haughk, T. Frauenheim, S. Suhai, and G. Seifert, Phys. Rev. B 
    {\bf 58}, 7260 (1996).

\bibitem{sanchez97} D. Sanches-Portal, P. Ordejon, E. Artacho, 
    and J. M. Soler, Int. J. Quantum Chem. {\bf 65}, 453 (1997).

\bibitem{langrethIJQC}
\bibinfo{author}{\bibfnamefont{D.~C.} \bibnamefont{Langreth}},
\bibinfo{author}{\bibfnamefont{M.}~\bibnamefont{Dion}},
\bibinfo{author}{\bibfnamefont{H.}~\bibnamefont{Rydberg}},
\bibinfo{author}{\bibfnamefont{E.}~\bibnamefont{Schr{\"o}der}},
\bibnamefont{and} \bibinfo{author}{\bibfnamefont{B.~I.}
\bibnamefont{Lundqvist}}, \bibinfo{journal}{Int. J. Quantum Chem.}
\textbf{\bibinfo{volume}{101}}, \bibinfo{pages}{599} (\bibinfo{year}{2005}).
         
\bibitem{aaron05} A. Puzder, M. Dion, and D. C. Langreth,
    submitted to J. Chem.~Phys.
    
\bibitem{svetla2}
\bibinfo{author}{\bibfnamefont{S.~D.} \bibnamefont{Chakarova-K\"ack}}
  \bibnamefont{and}
  \bibinfo{author}{\bibfnamefont{E.}~\bibnamefont{Schr\"oder}},
  \bibinfo{note}{to be published; this work supercedes an earlier
  method (see Ref.~\citen{svetla1} below) for extracting finite dimer energies from
  an infinite double layer}.

\bibitem{svetla1}
\bibinfo{author}{\bibfnamefont{S.~D.} \bibnamefont{Chakarova}}
\bibnamefont{and}
\bibinfo{author}{\bibfnamefont{E.}~\bibnamefont{Schr\"oder}},
\bibinfo{journal}{J. Chem.\ Phys.} \textbf{\bibinfo{volume}{122}},
\bibinfo{pages}{054102} (\bibinfo{year}{2005}).   
  
\bibitem{tsuzuki-naphthalene}
\bibinfo{author}{\bibfnamefont{S.}~\bibnamefont{Tsuzuki}},
\bibinfo{author}{\bibfnamefont{K.}~\bibnamefont{Honda}},
\bibinfo{author}{\bibfnamefont{T.}~\bibnamefont{Uchimaru}}, \bibnamefont{and}
\bibinfo{author}{\bibfnamefont{M.}~\bibnamefont{Mikami}},
\bibinfo{journal}{J. Chem.\ Phys.} \textbf{\bibinfo{volume}{120}},
\bibinfo{pages}{647} (\bibinfo{year}{2004}).

\bibitem{benz-graph}
\bibinfo{note}{S. D. Chakarova-K\"ack et al., to be published}.

\bibitem{pe}
\bibinfo{note}{J. Kleis et al., to be published}.  

\bibitem{DFT} P. Hohenberg and W. Kohn, 
    Phys. Rev. {\bf 136}, B864 (1964); 
    W. Kohn and L. J. Sham, Phys. Rev. {\bf 140}, A1133 (1965);

\bibitem{revPBE} Y. Zhang and W. Yang, 
    Phys. Rev. Lett. {\bf 80}, 890 (1998).

\bibitem{LDA} J. P. Perdew and Y. Wang, 
    Phys. Rev. B {\bf 45}, 13244 (1992).

\bibitem{gonze02} X. Gonze \emph{et al.}, Comp. Mat. Sci. {\bf 25},
    478 (2002).
    
\bibitem{troullier93} N. Troullier, J. L. Martins, Phys. Rev. 
    B {\bf 43}, 1993 (1991).
    
\bibitem{misquitta}
\bibinfo{author}{\bibfnamefont{A.~J.} \bibnamefont{Misquitta}},
\bibinfo{author}{\bibfnamefont{B.}~\bibnamefont{Jeziorski}},
\bibnamefont{and}
\bibinfo{author}{\bibfnamefont{K.}~\bibnamefont{Szalewicz}},
\bibinfo{journal}{Phys.\ Rev.\ Lett.} \textbf{\bibinfo{volume}{91}},
\bibinfo{pages}{033201} (\bibinfo{year}{2003}).

\bibitem{szal1}
\bibinfo{author}{\bibfnamefont{R.}~\bibnamefont{Podeszwa}} \bibnamefont{and}
\bibinfo{author}{\bibfnamefont{K.}~\bibnamefont{Szalewicz}},
\bibinfo{note}{preprint}.

\bibitem{szal2}
\bibinfo{author}{\bibfnamefont{K.}~\bibnamefont{Szalewicz}},
\bibinfo{note}{private communication}.  
    
\bibitem{lorentz}
 H. A. Lorentz,  \textit{Theory of Electrons}, 2nd edition, (1915), (Dover, New York, 1952).
 
\bibitem{jackson}
 J. D. Jackson, \textit{Classical Electrodynamics} (Wiley, New York, 1999).
 
 \bibitem{gamess} M. W. Schmidt \emph{et al.},
    J. Comput. Chem. {\bf 14}, 1347 (1993).

\bibitem{paliwal94} S. Paliwal, S. Geib, and C. S. Wilcox, J. Am. 
    Chem. Soc. {\bf 116}, 4497 (1994).

\bibitem{rashkin02} M. J. Rashkin and M. L. Waters, J. Am. Chem. Soc.
    {\bf 124}, 1860 (2002).

\bibitem{quad} E. J. Meijer and M. Sprik, J. Chem. Phys. {\bf 105},
    8684 (1996).
    
\bibitem{hunter90} C. A. Hunter and J. K. M. Sanders, J. Am. Chem. Soc.
    {\bf 112}, 5525 (1990).

\end{thebibliography}
\end{document}